\documentclass{PoS}

\usepackage{amsmath}
\usepackage{url}
\usepackage{ifthen} % for conditional statements
\newboolean{uprightparticles}
\setboolean{uprightparticles}{true} %Set to false to get italic particle symbols
\usepackage{xspace}     
\usepackage{amsfonts}
\usepackage{upgreek}

\newboolean{pdflatex}
\setboolean{pdflatex}{true} % use this if using non-eps figures

\def\ux85 {UX85\xspace}

\ifthenelse{\boolean{uprightparticles}}%
{

 \def\Pmu         {\ensuremath{\upmu}\xspace}

 \def\Ppi         {\ensuremath{\uppi}\xspace}

 \def\PDelta      {\ensuremath{\Delta}\xspace}                 
 \def\PXi      {\ensuremath{\Xi}\xspace}                 
 \def\PLambda      {\ensuremath{\Lambda}\xspace}                 
 \def\PSigma      {\ensuremath{\Sigma}\xspace}                 
 \def\POmega      {\ensuremath{\Omega}\xspace}                 
 \def\PUpsilon      {\ensuremath{\Upsilon}\xspace}                 
 
 \def\PB      {\ensuremath{\mathrm{B}}\xspace}                 
                  
 \def\PD      {\ensuremath{\mathrm{D}}\xspace}

 \def\PK      {\ensuremath{\mathrm{K}}\xspace}

 \def\Pb      {\ensuremath{\mathrm{b}}\xspace}

 \def\Ph      {\ensuremath{\mathrm{h}}\xspace}                 
 \def\Pi      {\ensuremath{\mathrm{i}}\xspace}

 \def\Pp      {\ensuremath{\mathrm{p}}\xspace}

}
{

 \def\Pmu         {\ensuremath{\mu}\xspace}

 \def\Ppi         {\ensuremath{\pi}\xspace}

 \mathchardef\PDelta="7101
 \mathchardef\PXi="7104
 \mathchardef\PLambda="7103
 \mathchardef\PSigma="7106
 \mathchardef\POmega="710A
 \mathchardef\PUpsilon="7107
                  
 \def\PB      {\ensuremath{B}\xspace}                 
                  
 \def\PD      {\ensuremath{D}\xspace}

 \def\PK      {\ensuremath{K}\xspace}

 \def\Pb      {\ensuremath{b}\xspace}

 \def\Ph      {\ensuremath{h}\xspace}                 
 \def\Pi      {\ensuremath{i}\xspace}

 \def\Pp      {\ensuremath{p}\xspace}

}

\def\mun        {\ensuremath{\Pmu^-}\xspace} % muon negative (\mum is taken)

\def\c     {\ensuremath{\Pc}\xspace}
\def\cbar  {\ensuremath{\overline \c}\xspace}

\def\b     {\ensuremath{\Pb}\xspace}

\def\pion  {\ensuremath{\Ppi}\xspace}

\def\pip   {\ensuremath{\pion^+}\xspace}
\def\pim   {\ensuremath{\pion^-}\xspace}
\def\pipi  {\ensuremath{\pion^+\pion^-}\xspace}

\def\kaon  {\ensuremath{\PK}\xspace}
  \def\Kbar  {\kern 0.2em\overline{\kern -0.2em \PK}{}\xspace}

\def\Kz    {\ensuremath{\kaon^0}\xspace}
\def\Kzb   {\ensuremath{\Kbar^0}\xspace}
\def\KzKzb {\ensuremath{\Kz \kern -0.16em \Kzb}\xspace}
\def\Kp    {\ensuremath{\kaon^+}\xspace}
\def\Km    {\ensuremath{\kaon^-}\xspace}

\def\KpKm  {\ensuremath{\Kp \kern -0.16em \Km}\xspace}
\def\KS    {\ensuremath{\kaon^0_{\rm\scriptscriptstyle S}}\xspace}

\def\Dbar    {\kern 0.2em\overline{\kern -0.2em \PD}{}\xspace}
\def\D       {\ensuremath{\PD}\xspace}

\def\Dz      {\ensuremath{\D^0}\xspace}
\def\Dzb     {\ensuremath{\Dbar^0}\xspace}
\def\DzDzb   {\ensuremath{\Dz {\kern -0.16em \Dzb}}\xspace}
\def\Dp      {\ensuremath{\D^+}\xspace}
\def\Dm      {\ensuremath{\D^-}\xspace}

\def\DpDm    {\ensuremath{\Dp {\kern -0.16em \Dm}}\xspace}

\def\Dstarp  {\ensuremath{\D^{*+}}\xspace}

  \def\Bbar    {\kern 0.18em\overline{\kern -0.18em \PB}{}\xspace}

  \def\Y#1S{\ensuremath{\PUpsilon{(#1S)}}\xspace}% no space before {...}!

\def\proton      {\ensuremath{\Pp}\xspace}

\newcommand{\decay}[2]{\ensuremath{#1\!\to #2}\xspace}         % {\Pa}{\Pb \Pc}

\def\to                 {\ensuremath{\rightarrow}\xspace}

\def\order   {\ensuremath{\mathcal{O}}\xspace}

\def\CP                {\ensuremath{C\!P}\xspace}

\def\AT#1     {\ensuremath{A_T^{#1}}\xspace}           % 2

\def\C#1      {\ensuremath{\mathcal{C}_{#1}}\xspace}                       % 9
\def\Cp#1     {\ensuremath{\mathcal{C}_{#1}^{'}}\xspace}                    % 7
\def\Ceff#1   {\ensuremath{\mathcal{C}_{#1}^{\mathrm{(eff)}}}\xspace}        % 9  
\def\Cpeff#1  {\ensuremath{\mathcal{C}_{#1}^{'\mathrm{(eff)}}}\xspace}       % 7
\def\Ope#1    {\ensuremath{\mathcal{O}_{#1}}\xspace}                       % 2
\def\Opep#1   {\ensuremath{\mathcal{O}_{#1}^{'}}\xspace}                    % 7

\def\agamma     {\ensuremath{A_{\Gamma}}\xspace}

\newcommand{\bra}[1]{\ensuremath{\langle #1|}}             % {a}
\newcommand{\ket}[1]{\ensuremath{|#1\rangle}}              % {b}
\newcommand{\tev}{\ensuremath{\mathrm{\,Te\kern -0.1em V}}\xspace}
\newcommand{\gev}{\ensuremath{\mathrm{\,Ge\kern -0.1em V}}\xspace}
\newcommand{\mev}{\ensuremath{\mathrm{\,Me\kern -0.1em V}}\xspace}
\newcommand{\kev}{\ensuremath{\mathrm{\,ke\kern -0.1em V}}\xspace}
\newcommand{\ev}{\ensuremath{\mathrm{\,e\kern -0.1em V}}\xspace}
\newcommand{\gevc}{\ensuremath{{\mathrm{\,Ge\kern -0.1em V\!/}c}}\xspace}
\newcommand{\mevc}{\ensuremath{{\mathrm{\,Me\kern -0.1em V\!/}c}}\xspace}
\newcommand{\gevcc}{\ensuremath{{\mathrm{\,Ge\kern -0.1em V\!/}c^2}}\xspace}
\newcommand{\gevgevcccc}{\ensuremath{{\mathrm{\,Ge\kern -0.1em V^2\!/}c^4}}\xspace}
\newcommand{\mevcc}{\ensuremath{{\mathrm{\,Me\kern -0.1em V\!/}c^2}}\xspace}

\def\mum  {\ensuremath{\,\upmu\rm m}\xspace}

\def\invfb   {\ensuremath{\mbox{\,fb}^{-1}}\xspace}

\def\order{{\ensuremath{\cal O}}\xspace}

\def\gsim{{~\raise.15em\hbox{$>$}\kern-.85em
          \lower.35em\hbox{$\sim$}~}\xspace}
\def\lsim{{~\raise.15em\hbox{$<$}\kern-.85em
          \lower.35em\hbox{$\sim$}~}\xspace}

\def\sqs   {\ensuremath{\protect\sqrt{s}}\xspace}

\def\pt         {\mbox{$p_T$}\xspace}

\def\tell1  {TELL1\xspace}
\def\ukl1   {UKL1\xspace}

\def\deltam     {\ensuremath{\Delta m}\xspace}

\def\hphm       {\ensuremath{\Ph^+ \Ph^-}\xspace}

\def\gammahat   {\ensuremath{\hat{\Gamma}}\xspace}

\def\agammadefnot  {\ensuremath{\frac{\gammahat(\Dz \to f) - \gammahat(\Dzb \to f)}{\gammahat(\Dz \to f) + \gammahat(\Dzb \to f)}}\xspace}
\def\etacp      {\ensuremath{\eta_{\CP}}\space}

\def\dzkk       {\decay{\Dz}{\Kp\Km}}
\def\dzpipi     {\decay{\Dz}{\pip\pim}}

\def\dzkpi      {\decay{\Dz}{\Km\pip}}

\def\dstdzpi    {\decay{\Dstarp}{\Dz\pip_s}\xspace}

\newcommand{\sqseq}[1]{\ensuremath{\sqs =\text{#1}}\xspace}

\def\c          {\ensuremath{c}\xspace}

\newcommand{\magnitude}[1]{\ensuremath{\left|{#1}\right|}\xspace}
\newcommand{\magsq}[1]{\ensuremath{\magnitude{#1}^2}\xspace}

\newcommand{\orderof}[1]{\ensuremath{\order(#1)}\xspace}

\newcommand{\tene}[1]{\ensuremath{10^{#1}}\xspace}
\newcommand{\xtene}[2]{\ensuremath{#1 \times \tene{#2}}\xspace}
\newcommand{\otene}[1]{\orderof{\tene{#1}}}

\def\hamiltonian{\ensuremath{\mathcal{H}}\xspace}

\def\C	{\ensuremath{C}\xspace}

\newcommand{\sqrtseq}[1]{\ensuremath{\sqs = #1}\xspace}

\newcommand{\cpasym}[2]{\ensuremath{\frac{#1-#2}{#1+#2}}\xspace}

\def\Abar{\bar{A}}
\def\fbar{\bar{f}}

\def\Af{\ensuremath{A_f}\xspace}
\def\Afbar{\ensuremath{A_{\fbar}}\xspace}
\def\Abarf{\ensuremath{\Abar_f}\xspace}
\def\Abarfbar{\ensuremath{\Abar_{\fbar}}\xspace}

\newcommand{\amplitudedef}[2]{\ensuremath{\bra{#1} \hamiltonian \ket{#2}}\xspace}
\newcommand{\Afdef}[1]{\amplitudedef{f}{#1}}
\newcommand{\Afbardef}[1]{\amplitudedef{\fbar}{#1}}

\def\Addef{\cpasym{\magsq{\Af}}{\magsq{\Abarfbar}}}

\def\DHLDef{\ensuremath{\ket{\PD_{L,H}} = p \ket{\Dz} \pm q \ket{\Dzb}}\xspace}
\def\xmixdef{\ensuremath{\frac{2(m_H - m_L)}{\Gamma_H + \Gamma_L}}\xspace}
\def\ymixdef{\ensuremath{\frac{\Gamma_H - \Gamma_L}{\Gamma_H + \Gamma_L}}\xspace}

\def\Lambdaf{\ensuremath{\lambda_f}\xspace}
\def\Lambdafdef{\ensuremath{\frac{q \Af}{p \Abarf}}\xspace}
\def\Lambdafmagdef{\ensuremath{\magnitude{\Lambdafdef} e^{i\phi}}\xspace}

\def\pis{\ensuremath{\pip_s}\xspace}

\def\btodzmux{\decay{\PB}{\Dz \mun X}}

\graphicspath{
  {figs/}
}

\def\dzksks{\decay{\Dz}{\KS\KS}}
\def\dzhphm{\decay{\Dz}{\hphm}}

\title{Searches for CP violation in two-body charm decays}

\ShortTitle{Searches for CP violation in two-body charm decays}

\author{\speaker{Michael Alexander}%
  \thanks{On behalf of the LHCb collaboration.}\\
  University of Glasgow (GB)\\
  E-mail: \email{michael.alexander@glasgow.ac.uk}}

\abstract{
  The LHCb experiment recorded data corresponding to an integrated luminosity of 3.0 \invfb during its first run of data taking. These data yield the largest samples of charmed hadrons in the world and are used to search for \CP violation in the \Dz system. Among the many measurements performed at LHCb, a measurement of the direct \CP asymmetry in \dzksks decays is presented and is found to be
\begin{align*}
  A_{CP}(\dzksks) = (-2.9 \pm 5.2 \pm 2.2)\, \%,
\end{align*}
where the first uncertainty is statistical and the second systematic. This represents a significant improvement in precision over the previous measurement of this parameter. Measurements of the parameter \agamma, defined as the \CP asymmetry of the \Dz effective lifetime when decaying to a \CP eigenstate, are also presented. Using semi-leptonic \b-hadron decays to tag the flavour of the \Dz meson at production with the \KpKm and \pipi final states yields
\begin{align*}
  \agamma(\KpKm) &= (-0.134 \pm 0.077^{+0.026}_{-0.034})\, \%, \nonumber \\
  \agamma(\pipi) &= (-0.092 \pm 0.145^{+0.025}_{-0.033})\, \%.
\end{align*}
Thus no evidence of direct or indirect \CP violation in the \Dz system is found, though it is tightly constrained. 
}

\FullConference{The European Physical Society Conference on High Energy Physics\\
		 22-29 July 2015\\
		 Vienna, Austria}

\begin{document}

\section{Introduction}

The \Dz meson is the only heavy, neutral meson comprised from up-type quarks in the Standard Model (SM) making it a unique system in which to study \CP violation. Due to the form of the CKM matrix \CP violation in interactions involving charm quarks is strongly suppressed. For a final state $f$, defining the amplitudes
$\Af = \Afdef{\Dz}$, $\Afbar = \Afbardef{\Dz}$, $\Abarf = \Afdef{\Dzb}$, and $\Abarfbar = \Afbardef{\Dzb}$,
%\begin{align}
%  \Af &= \Afdef{\Dz}, & \Afbar &= \Afbardef{\Dz}, \nonumber \\
%  \Abarf &= \Afdef{\Dzb}, & \Abarfbar &= \Afbardef{\Dzb},
%\end{align}
with \hamiltonian the Hamiltonian, direct \CP violation is quantified by
\begin{align}
  A_{CP}^{dir} \equiv \Addef.
\end{align}
In the SM direct \CP violation in the decays \dzkk and \dzpipi is predicted to be up to \otene{-3}\cite{lhcb_implications2013}. 

As the \Dz meson is neutral it can oscillate into a \Dzb meson, and vice-versa. Consequently, the mass eigenstates in which the \Dz and \Dzb mesons propagate are superpositions of the flavour eigenstates, defined by
\begin{align}
  \DHLDef,
\end{align}
with masses $m_{H,L}$ and widths $\Gamma_{H,L}$. Here $p$ and $q$ are complex, satisfying $\magsq{p} + \magsq{q} = 1$. The rate of mixing is quantified by
\begin{align}
  x = \xmixdef,\,\,\mathrm{and}\,\, y = \ymixdef.
\end{align}
\Dz mixing is now firmly established experimentally though uncertainties are still relatively large \cite{lhcb_charmmixing2013}. \CP violation in mixing is quantified (following the conventions of \cite{HFAGApril2015}) by
\begin{align}
  A_{CP}^{mix} = \magsq{\frac{q}{p}} - 1.
\end{align}
For a final state accessible to both \Dz and \Dzb mesons \CP violation can arise from interference between mixing and decay, which is quantified by
\begin{align}
  \Lambdaf \equiv \Lambdafdef = \Lambdafmagdef.
\end{align}
Such indirect \CP violation is predicted to be up to \otene{-4} in the SM \cite{Bobrowski_indirectCPVCharm2010}. Observation of larger direct or indirect \CP violation than SM predictions would be a strong indication of new physics.

The LHCb detector at the LHC, CERN, is a forward arm spectrometer covering the high pseudo-rapidity region $2 < \eta < 5$ and is specifically designed to perform high precision measurements of decays involving $b$ and $c$ quarks \cite{JINST_LHCb}. Key components of the detector are: the Vertex Locator (VELO), which provides fine tracking around the interaction point and achieves impact parameter resolutions of \mbox{$\sim$20 \mum} for tracks with \pt $>$ 1 \gev \cite{lhcb_veloPerformance2014}; two Ring Imaging Cherenkov detectors providing particle identification with excellent separation of \pion and \kaon mesons across a wide momentum range \cite{lhcb_richPerformance2014}; and the tracking stations, positioned before and after the dipole magnet, which achieve momentum resolutions of $\sim$0.5-0.8 \% \cite{lhcb_detectorPerformance2014}.

During its first data-taking run the LHCb experiment recorded data corresponding to integrated luminosities of 1.0 \invfb at \sqrtseq{7 \tev} in 2011 and 2.0 \invfb at \sqrtseq{8 \tev} in 2012. As the $\c$$\cbar$ production cross section in the $\proton$$\proton$ collisions provided by the LHC is very large \cite{lhcb_promptCharmProduction2013} this has yielded the largest data sets of charm meson decays in the world. This has allowed the LHCb experiment to perform many of the highest precision measurements of \CP violation in the \Dz system to date. Among these are the measurement of direct \CP violation in \dzksks decays \cite{lhcb_D0KsKs2015}, presented in Sec. \ref{sec:ksks}, and the measurements of indirect \CP violation in \dzkk and \dzpipi \cite{lhcb_SLAGamma2015}, presented in Sec. \ref{sec:slagamma}.

\section{Direct \CP violation in {$\boldsymbol{\dzksks}$}}
\label{sec:ksks}

The dominant, tree-level amplitudes in \dzksks decays largely cancel, meaning the $\KS\KS$ final state is predominantly reached through final state scattering \decay{\dzpipi}{\KS\KS} and \decay{\dzkk}{\KS\KS}. The additional interference introduced by this re-scattering can enhance direct \CP violation to \otene{-2}. The only previous measurement of this found \mbox{$A_{CP}(\dzksks) = (23 \pm 19) \%$} \cite{CLEO_D0KsKs}.

To determine $A_{CP}(\dzksks)$, firstly \decay{\KS}{\pipi} candidates are reconstructed and combined to form the \Dz candidates. These are required to have invariant mass within $\pm 20$ \mev of the known \Dz mass \cite{PDG2015}. As this is a \CP eigenstate final state there is no detection asymmetry. The flavour of the \Dz candidates at production is determined using \dstdzpi decays, where the charge of the ``soft pion'', \pis, determines the \Dz flavour. Decays of \dstdzpi with \dzkpi are used to determine the \Dstarp production and \pis detection asymmetries. The background from \decay{\Dz}{\KS\pipi} decays is removed using a minimum flight distance cut on the \KS candidates. The remaining background is combinatorial which is minimised via a multi-variate selection using candidate kinematics, decay times, geometry and decay-tree fit quality variables.

Candidates are accepted if any track in the event, excluding that of the \pis candidate, has triggered the event. As \KS mesons are relatively long lived slightly more than half of their decays occur outside the acceptance of the VELO. Those which decay outside the VELO but before the tracking station in front of the dipole magnet (known as downstream candidates) are still reconstructed but with worse momentum and vertex resolution than those which decay within the VELO (known as long candidates). Consequently the \Dz candidates are divided into subsets where: both \KS candidates are long (LL); one \KS candidate is long and the other downstream (LD); both \KS candidates are downstream (DD). Additionally, a dedicated trigger for LL candidates was implemented for 2012 data taking giving a much cleaner data sample, so candidates satisfying this trigger (labelled LLtrig) are separated from other LL candidates.

\begin{figure}[t]
  \centering
  \includegraphics[width=0.4\textwidth]{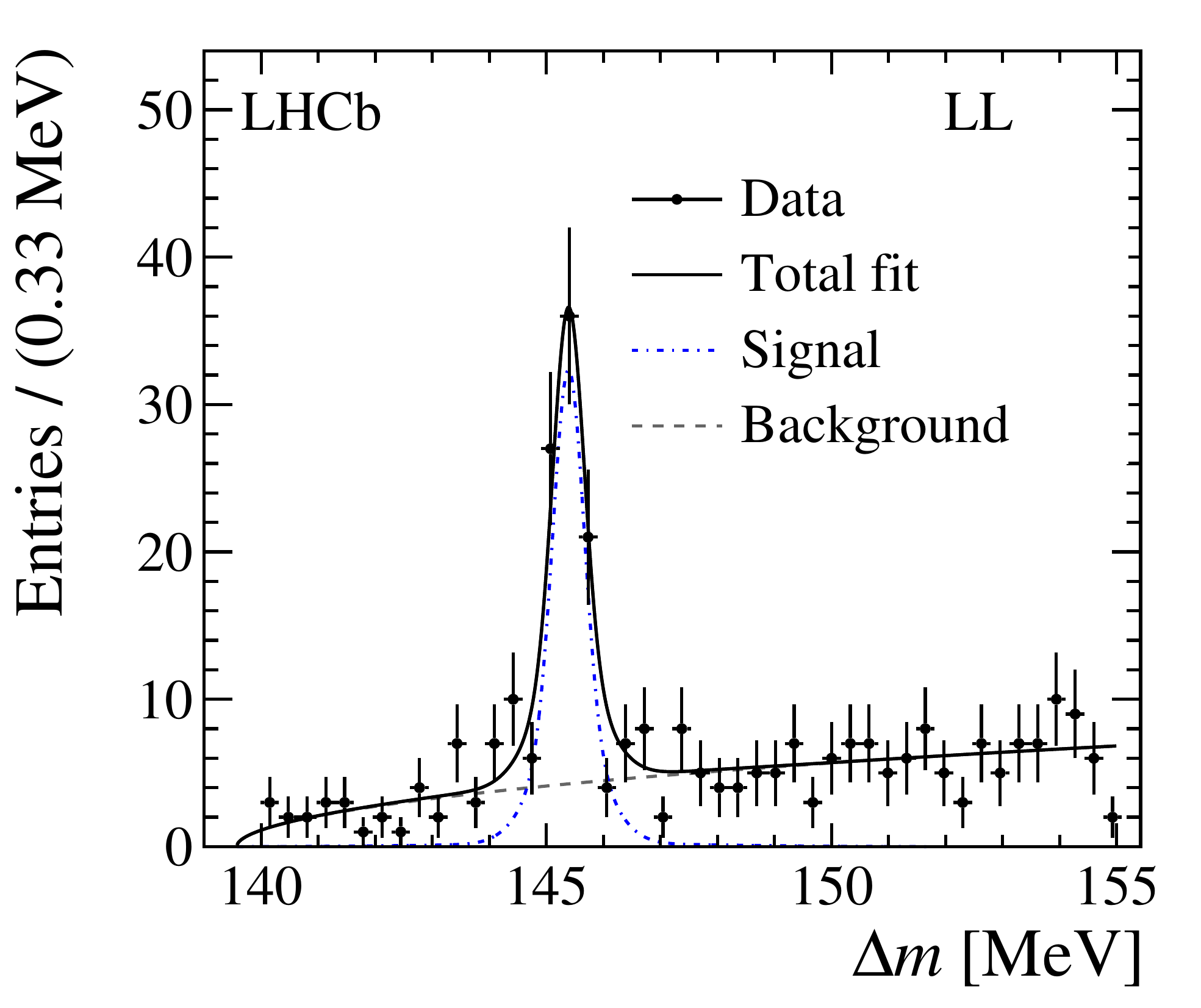}
  \includegraphics[width=0.4\textwidth]{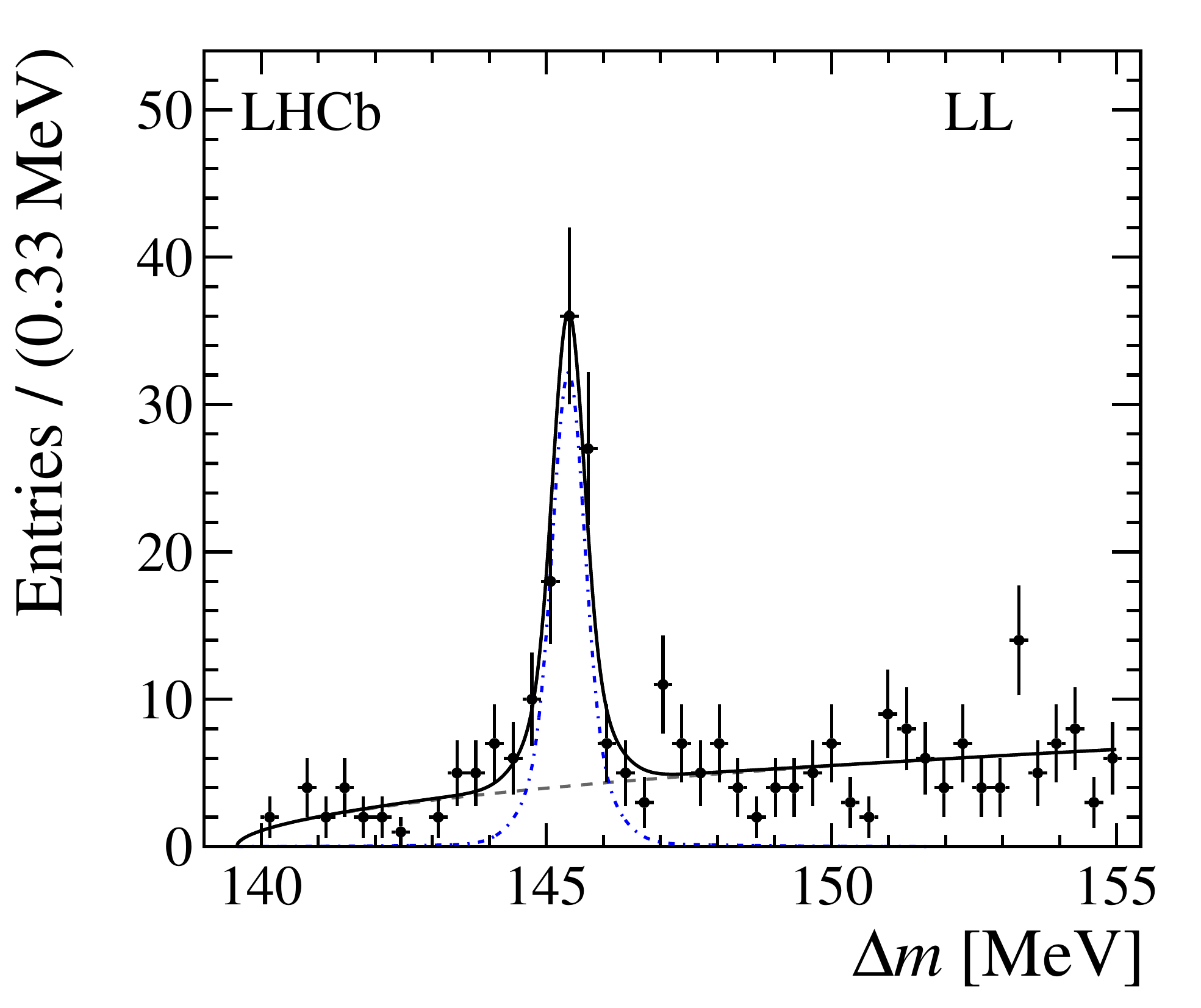}
  \caption[\deltam fits for LL \dzksks candidates]{Fits to the distribution of \deltam for (left) \Dz and (right) \Dzb LL candidates.
    %The solid black line is the total fit function, the dashed grey line is the background, and the dot-dashed blue line the signal.
  }
  \label{fig:KsKsDmLL}
\end{figure}

To improve the \Dz and \Dstarp mass resolutions the mass of the \KS candidates is constrained to the known \KS mass and the trajectory of the \Dz candidate is constrained to originate from primary interaction point. The yields of \Dz and \Dzb signal are then determined via fits to the distribution of $\deltam \equiv m(\Dstarp) - m(\Dz)$. The signal is modelled using a sum of three Gaussians and the combinatorial background with an empirical threshold function. The four subsets (LLtrig, LL, LD, DD) are fitted independently. For each fit the shape parameters for signal and background are shared between \Dz and \Dzb candidates. Fig. \ref{fig:KsKsDmLL} shows the \deltam fits for LL candidates. In total, approximately 650 signal decays are found. 

The final result is obtained by taking a simple weighted average of the asymmetries of each subset and is found to be
\begin{align}
  A_{CP}(\dzksks) = (-2.9 \pm 5.2 \pm 2.2)\, \%,
\end{align}
where the first uncertainty is statistical and the second systematic. The dominant systematic uncertainties arise from the accuracy of the fit model and the \Dstarp production and \pis detection asymmetries (determined from the control channel). This represents a significant improvement in precision over the previous measurement, but shows no indication of \CP violation. For the second data taking run the precision achieved is expected to improve significantly due to the implementation of dedicated triggers for LD and DD candidates. 

\section{Indirect \CP violation in semi-leptonic-tagged {$\boldsymbol{\dzhphm}$}}
\label{sec:slagamma}

The \CP asymmetry of the effective lifetime of the \Dz meson decaying to a \CP eigenstate final state, $f$, is primarily sensitive to indirect \CP violation as \cite{aGammaYCPTheory}
\begin{align}
  \agamma \equiv \agammadefnot \approx \etacp \left[ \left(A^{mix}_{CP}/2 - A^{dir}_{CP}\right) y \cos \phi - x \sin \phi \right],
\end{align}
where \gammahat is the inverse of the effective lifetime and \etacp is the \CP eigenvalue of $f$. For \dzpipi and \dzkk decays $\etacp = 1$. Previous measurements of \agamma using \Dstarp tagged \Dz decays have found no evidence of \CP violation \cite{lhcb:AGamma2013}.

A complementary flavour tagging technique is to use \btodzmux decays, where the charge of the \mun gives the flavour of the \Dz at production and $X$ denotes any other decay products of the \PB that aren't considered. To measure \agamma, combinatorial backgrounds are firstly minimised by applying cuts to kinematic and decay-tree fit quality variables. The yields of \Dz and \Dzb are obtained from fits to the distributions $m(\Dz)$ in bins of \Dz decay time in order to calculate
\begin{align}
  A_{CP}(t) \simeq A^{dir}_{CP} - \agamma \frac{t}{\tau},
\end{align}
where $\tau$ is the world average of the effective lifetime of the \Dz meson in \dzhphm decays. The measured value of $A^{dir}_{CP}$ includes the \PB production and \mun detection asymmetries. The reconstruction efficiency as a function of decay time cancels in the asymmetry calculation and, assuming no asymmetry in the mistag rate, any mistagging only reduces the sensitivity to \agamma without introducing a bias.

\begin{figure}[t]
  \centering
  \includegraphics[width=0.4\textwidth]{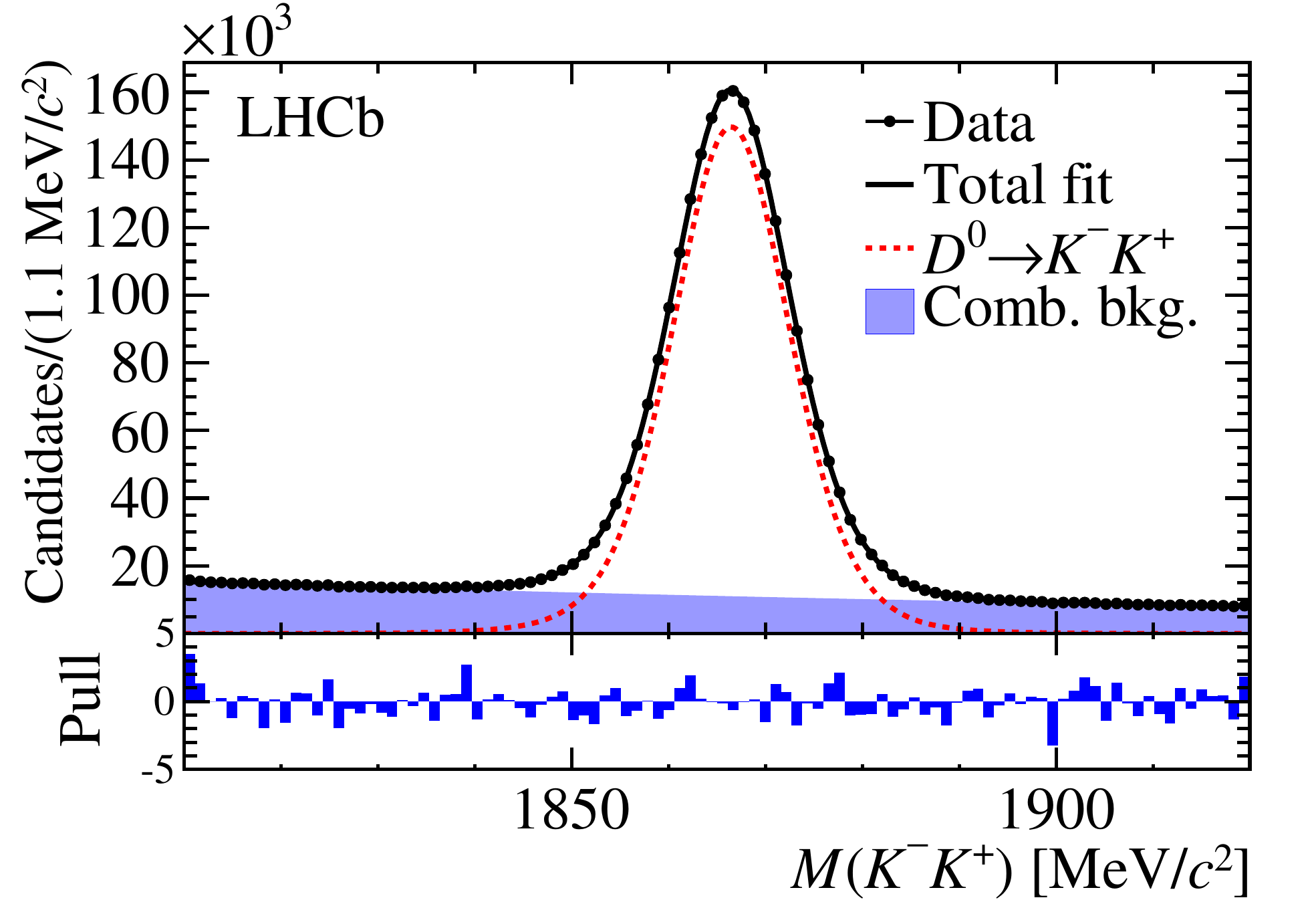}
  \includegraphics[width=0.4\textwidth]{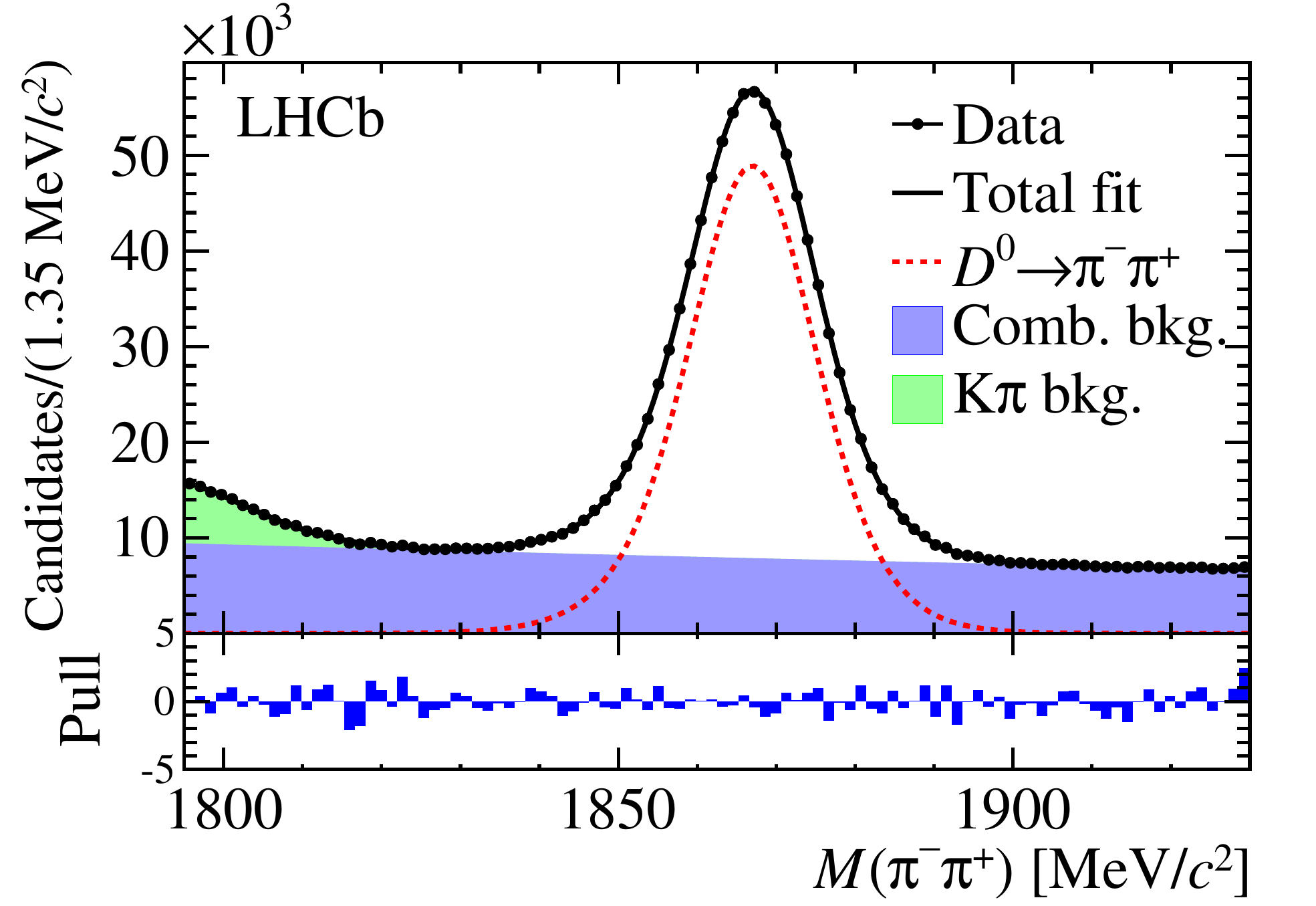}
  \caption[\dzkk and \dzpipi mass fits]{Decay-time integrated fits to the $m(\Dz)$ distributions for (left) \dzkk and (right) \dzpipi candidates.}
  \label{fig:SLAGammaMass}
\end{figure}

\begin{figure}[t]
  \centering
  \includegraphics[width=0.4\textwidth]{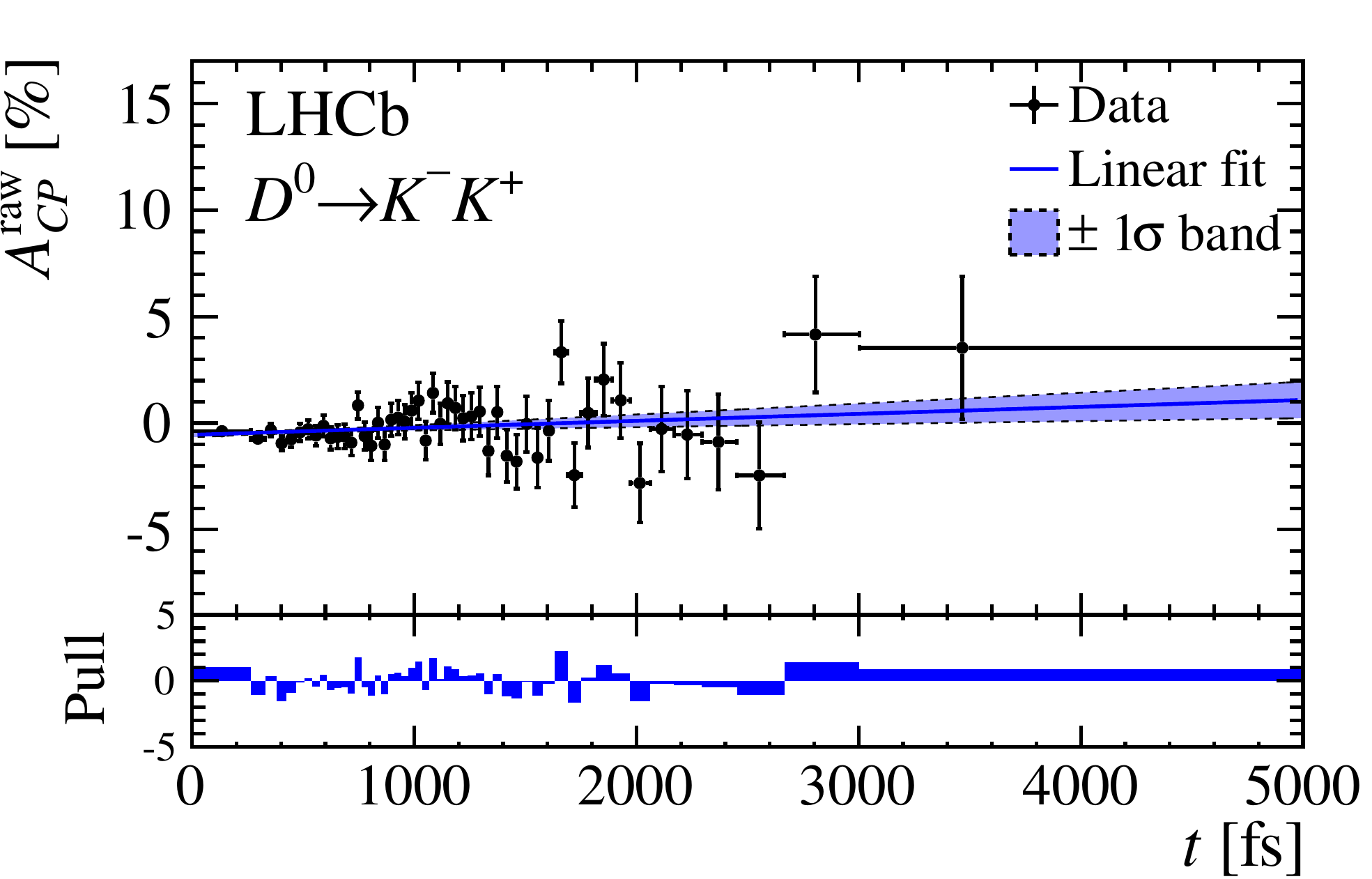}
  \includegraphics[width=0.4\textwidth]{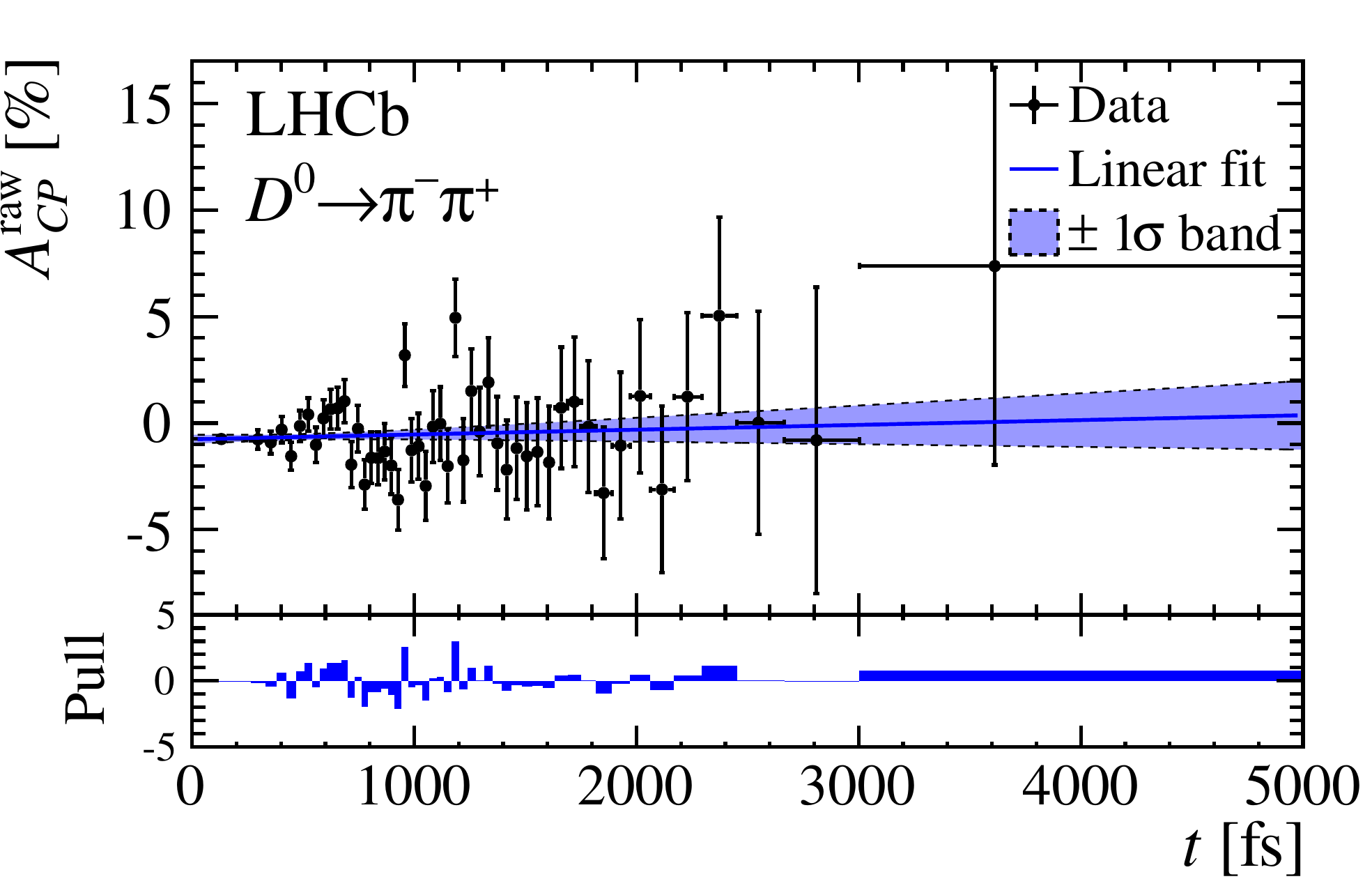}
  \caption[$A_{CP}(t)$ for \dzkk and \dzpipi]{\CP asymmetry as a function of decay time for (left) \dzkk and (right) \dzpipi candidates.}
  \label{fig:SLAGammaACPt}
\end{figure}

Fig. \ref{fig:SLAGammaMass} shows the decay-time integrated fits to the $m(\Dz)$ distributions for \dzkk and \dzpipi candidates. The shape parameters are fixed from these fits for the fits in bins of decay time. \Dz and \Dzb candidates are fitted simultaneously in order to determine the asymmetry directly. Approximately \xtene{2.34}{6} \dzkk and \xtene{0.79}{6} \dzpipi signal decays are found. Fig. \ref{fig:SLAGammaACPt} shows the fits to the \CP asymmetries as a function of decay time, which find
\begin{align}
  \agamma(\KpKm) &= (-0.134 \pm 0.077^{+0.026}_{-0.034})\, \%, \nonumber \\
  \agamma(\pipi) &= (-0.092 \pm 0.145^{+0.025}_{-0.033})\, \%,
\end{align}
where the first uncertainty is statistical and the second systematic. The sources of systematics are well understood from the \dzkpi control channel and are expected to scale with statistics for the second data taking run.

These measurements show no indication of \CP violation. Including them in the world average gives a result of $-\agamma \simeq a_{CP}^{ind} = (0.058 \pm 0.040)\,\%$ \cite{HFAGApril2015}. Combining this with measurements of direct \CP violation in \dzhphm decays in a global fit yields a p-value for \CP conservation of 1.8 \%. 

\section{Conclusions}
\label{sec:conclusions}

During the first data taking run the LHCb experiment recorded data corresponding to an integrated luminosity of 3.0 \invfb. These data have been used to measure the direct \CP asymmetry in \dzksks decays, finding $A_{CP}(\dzksks) = (-2.9 \pm 5.2 \pm 2.2)\, \%$. This represents a significant improvement over the previous measurement. Indirect \CP asymmetries in \dzhphm decays have also been measured using semi-leptonic \PB decays to tag the \Dz flavour at production, finding $\agamma(\KpKm) = (-0.134 \pm 0.077^{+0.026}_{-0.034})\, \%$ and $\agamma(\pipi) = (-0.092 \pm 0.145^{+0.025}_{-0.033})\, \%$. Thus \CP violation in the \Dz system remains consistent with zero, though it is tightly constrained. For the second data taking run the precision on $A_{CP}(\dzksks)$ is expected to improve considerably due to the use of additional dedicated trigger lines. Measurements of \agamma will also benefit from improvements in triggering in addition to increased production cross sections at \sqseq{13 \tev}. The third data taking run, following the LHCb detector upgrade, will yield another order of magnitude in statistics. Thus the potential is high for the discovery of \CP violation in the \Dz system, and perhaps new physics, at the LHCb experiment in the coming years.

\bibliographystyle{unsrt}
\bibliography{bibliography.bib}

\end{document}